\documentclass[12pt,a4paper,twoside]{article}
\usepackage[latin1]{inputenc}
\usepackage[english]{babel}
\usepackage[margin=3cm]{geometry}

\usepackage{hyperref}
\usepackage{graphicx}
\usepackage{times}
\usepackage{fancyhdr}
\usepackage{epsfig}
\usepackage{multirow}
\usepackage{amssymb,amsfonts,amsmath}
\usepackage{cite}

\usepackage[hang,small,bf]{caption}

\newcommand{\panino}[1]{\left\langle #1\right\rangle}

\newcommand{\mca}[1]{{\mathcal #1}}
\newcommand{\de}[1]{\!\operatorname{d}\!{#1}}
\newcommand{\dec}[1]{\!\!\!\!\!\!\operatorname{d}\!{#1}}

\begin{document}

\thispagestyle{empty}

\begin{center}
 {\Large\bfseries\sffamily
 Epidemic Threshold in\\Continuous-Time Evolving Networks\\
 }
 \vspace{1cm}\large
{Eugenio Valdano\textsuperscript{1,$\dagger$},
Michele Re Fiorentin\textsuperscript{2},
Chiara Poletto\textsuperscript{1},\\
and Vittoria Colizza\textsuperscript{1,3}}\\
\vspace{1.1cm}
{\footnotesize
\textsuperscript{1}Sorbonne Universit\'es, UPMC Univ Paris 06, INSERM, Institut Pierre Louis d'\'Epid\'emiologie et de Sant\'e Publique (IPLESP~UMRS 1136), F75012, Paris, France.
\textsuperscript{2}Center for Sustainable Future Technologies, CSFT@PoliTo, Istituto Italiano di Tecnologia, corso Trento 21, 10129 Torino, Italy.
\textsuperscript{3}ISI Foundation, Torino, Italy.
\textsuperscript{$\dagger$}Current affiliation: Department d'Enginyeria Inform\`atica i Matem\`atiques, Universitat Rovira i Virgili, 43007 Tarragona, Spain.\\
}
\end{center}

 \vspace{1cm}

{\sffamily\footnotesize
\centering{\bfseries Abstract}\\
Current understanding of the critical outbreak condition on temporal networks relies on approximations (time scale separation, discretization) that may bias the results. We propose a theoretical framework to compute the epidemic threshold in continuous time through the infection propagator approach. We introduce the {\em weak commutation} condition allowing the interpretation of  annealed networks, activity-driven networks, and time scale separation into one formalism. Our work provides a coherent connection between discrete and continuous time representations applicable to realistic scenarios.
}

\vspace{1.5cm}



Contagion processes, such as the spread of diseases, information, or innovations~\cite{Keeling2007,Goffman1964,Daley1964,PastorSatorras2001,Watts2002}, share a common theoretical framework coupling the underlying population contact structure  with contagion features to provide an understanding of the resulting spectrum of  emerging collective behaviors~\cite{Vespignani2012}. A common keystone property  is the presence of a {threshold behavior} defining the transition between a macroscopic-level spreading regime and one characterized by a null or negligibly small contagion of individuals.   
Known as the {\it epidemic threshold} in the realm of infectious disease dynamics~\cite{Keeling2007}, the concept is analogous to the phase transition in non-equilibrium physical systems~\cite{Harris1974,Grassberger1983}, and is also central in social contagion processes~\cite{Watts2002,Centola2007,Liu2003,Moreno2004b,Ruan2015,Bottcher2017}.

A vast array of theoretical results characterize the epidemic threshold~\cite{PastorSatorras2015}, mainly under the limiting assumptions of quenched and annealed networks~\cite{Cohen2000,PastorSatorras2001,Newman2002,Wang2003,Gomez2010}, i.e., when the time scale of the network evolution is much slower or much faster, respectively, than the dynamical process. The recent availability of data on time-resolved contacts of epidemic relevance~\cite{Holme2015} has, however challenged the time scale separation, showing  it may introduce important biases in the description of the epidemic spread~\cite{Vazquez2007,Kao2007,Volz2009,Fefferman2007,Vernon2009,Iribarren2009,Karsai2011,Miritello2011,Stehle2011,Bajardi2011,Kivela2012,Rocha2013,Ferreri2014,Masuda2013,Holme2015} and in the characterization of the transition behavior~\cite{Bansal2010,Perra2012,Rocha2013,Starnini2014,Sun2015}. Departing from traditional approximations, few novel approaches are now available that derive the epidemic threshold constrained to specific contexts of generative models of temporal networks~\cite{Eames2004,Gross2006,Volz2009,Zhao2010,Perra2012,Taylor2012,Ferreri2014} or considering  generic discrete-time evolving contact patterns~\cite{Prakash2010,Valdano2015b,Valdano2015c}. In particular, the recently introduced infection propagator approach~\cite{Valdano2015b,Valdano2015c} is based on a matrix encoding the  probabilities of transmission of the infective agent along time-respecting paths in the network. Its spectrum allows the computation of the epidemic threshold at any given time scale and for an arbitrary discrete-time temporal network. Leveraging an original mapping of the temporal network and epidemic spread in terms of a multilayer structure, the approach is valid  in the discrete representation only, similarly to previous methods~\cite{Wang2003,Gomez2010,Perra2012}.

Meanwhile, a large interest in the study of continuously evolving temporal networks has developed, introducing novel representations~\cite{Boulicaut2004,Vazquez2007,Miritello2011,Holme2015} and proposing optimal discretization schemes~\cite{Krings2012,Holme2013,Valdano2015c} that may however be inaccurate close to the critical conditions ~\cite{Fennell2016}.
Most importantly, the two representations~--~continuous and discrete~--~of a temporal network remain disjointed in current network epidemiology. A discrete-time evolving network is indeed a multilayer object interpretable as a tensor in a linear algebraic representation~\cite{Dedomenico2013}. This is clearly no longer applicable when time is continuous, as it cannot be expressed in the form of successive layers. Hence, a coherent theoretical framework to bridge the gap between the two representations is still missing. 

In this Letter, we address this issue by
analytically deriving the infection propagator in continuous time. 
Formally, we show that the dichotomy discrete time~--~continuous time translates into the separation between a linear algebraic approach and a differential one, and that the latter can be derived as the structural limit of the former. Our approach yields a solution for the threshold of epidemics spreading on generic continuously evolving networks, and a closed form under a specific condition that is then validated through numerical simulations. 
In addition, the proposed novel perspective allows us to cast an important set of network classes into one single rigorous and  comprehensive mathematical definition, including annealed~\cite{Boguna2009,PastorSatorras2001,Castellano2010} and activity-driven~\cite{Perra2012,Liu2014} networks, widely used in both methodological and applied research.

\begin{figure}
\begin{center}
 \includegraphics[width=0.75\textwidth]{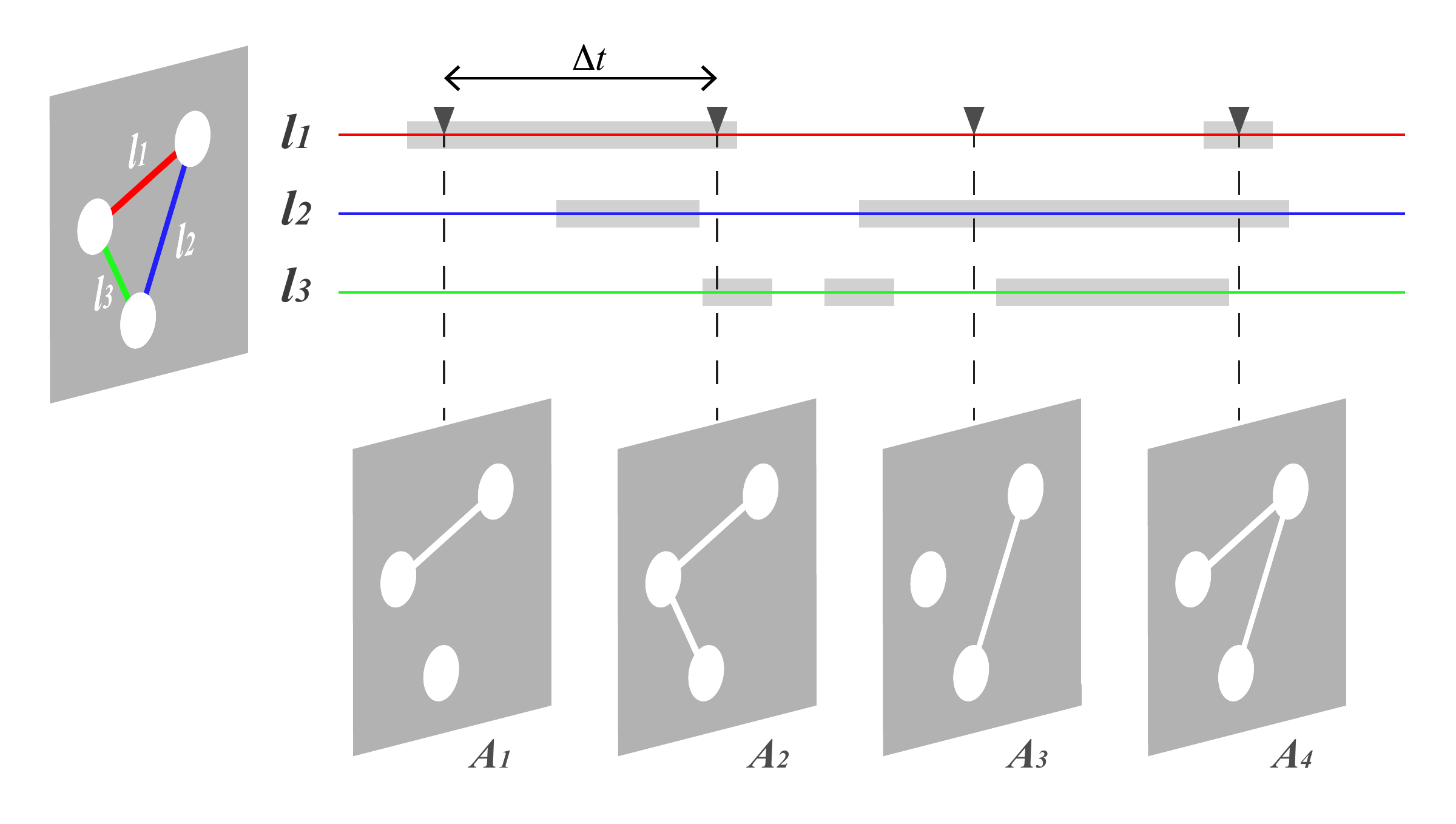}
\caption{\label{fig:sampling} Discrete sampling of a continuous-time temporal network. Links ($l_1$, $l_2$, $l_3$)  activate in time as marked by the colored segments (top). This time evolution is sampled at intervals $\Delta t$, building a sequence of snapshots (bottom), corresponding to adjacency matrices $\{A_1,A_2,\cdots\}$.
}
\end{center}
 \end{figure}

Let us consider a susceptible-infected-susceptible (SIS) epidemic model unfolding on a continuously evolving temporal network of $N$ nodes. The SIS model constitutes a basic paradigm for the description of epidemics with reinfection~\cite{Keeling2007}. Infectious individuals (I) can propagate the contagion to susceptible neighbors (S) with rate $\lambda$, and recover to the S state with rate $\mu$. The temporal network is described by the adjacency matrix $A(t)$, with $t\in\left[0,T\right]$.
We consider a discretized version of the system by sampling $A(t)$ at discrete time steps of length $\Delta t$ (Fig.~\ref{fig:sampling}). This yields a finite sequence of adjacency matrices $\{A_1,A_2,\cdots,A_{T_{step}}\}$, where $T_{step} = \left\lfloor T/\Delta t\right\rfloor$, and $A_h = A(h\Delta t)$. The sequence approximates the original continuous-time network with increasing accuracy as $\Delta t$ decreases. We describe the SIS dynamics on this discrete sequence of static networks as a discrete-time Markov chain~\cite{Wang2003,Gomez2010}:
\begin{align}
p_{h+1 , i} = & \left(1-p_{h , i}\right) [1-  \prod_j \left( 1-\lambda \Delta t  A_{h,ji}\; p_{h, j} \right) ] \nonumber \\
& + p_{h , i} \left(1-\mu \Delta t \right),
 \label{eq:temporal}
\end{align}
where $ p_{h,\, i}$ is the probability that a node $i$ is in the infectious state at  time step $h$, and $\mu\Delta t$ ($\lambda\Delta t$) is the probability that a node recovers (transmits the infection) during a time step $\Delta t$, for sufficiently small $\Delta t$.

By mapping the system into a multilayer structure encoding both network evolution and diffusion dynamics, the infection propagator approach derives the epidemic threshold as the solution of the equation $\rho\left(P(T_{step})\right)=1$~\cite{Valdano2015b,Valdano2015c}, where $\rho$ is the spectral radius of the following matrix:
\begin{equation}
P(T_{step})= \prod_{k=1}^{T_{step}} \left[ 1-\mu\Delta t + \lambda\Delta t A_k \right]\,.
\end{equation}
The generic element $P_{ij}(T_{step})$ represents the probability that the infection can propagate from node $i$ at time step $1$ to node $j$ at time step $T_{step}$, when $\lambda$ is close to $\lambda_c$ and within the quenched mean-field approximation (locally tree-like network~\cite{Radicchi2016}). For this reason, $P$ is denoted as the infection propagator. 

To compute the continuous-time limit of the infection propagator, we observe that $P$ obeys the recursive relation $P(h+1) = P(h)\left[ 1-\mu\Delta t + \lambda\Delta t A_{h+1} \right]$. Expressed in continuous time and dividing both sides by $\Delta t$, the relation becomes:
\begin{equation}
\frac{P(t+\Delta t) - P(t)}{\Delta t} =  P(t) \left[ -\mu + \lambda A(t+\Delta t) \right],
\label{eq:recursionFinal}
\end{equation}
that in the limit $\Delta t \rightarrow 0$ yields
\begin{equation}
\dot{P}(t) = P(t) \left[ -\mu+ \lambda A(t) \right]\,,
\label{eq:contDifferential}
\end{equation}
a system of $N^2$ coupled differential equations whose components are
\begin{equation}
\dot{P}_{ij}(t) = \sum_k \left[ \lambda A_{kj}(t) - \mu\delta_{kj}\right] P_{ik}(t).
\label{eq:contDifferentialComp}
\end{equation}
The {\itshape lhs} of Eqs.~(\ref{eq:contDifferential}) is the derivative of $P$ that is well behaved if all entries are continuous functions of time. $A_{ij}(t)$  are, however often binary, so that their evolution is a sequence of discontinuous steps. To overcome this, it is possible to approximate these steps with one-parameter families of continuous functions, compute the threshold, and then perform the limit of the parameter that recovers the discontinuity. More formally, this is equivalent to interpret derivatives in the sense of tempered distributions~\cite{Whittaker2000}.


In order to check that our limit process correctly connects the discrete-time framework to the continuous time one, let us now consider the standard Markov chain formulation of the continuous dynamics:
\begin{equation}
 \dot{p}_i(t) = \lambda \left[1-p_i(t)\right] \sum_j A_{ij}(t) p_j(t) - \mu p_i(t).
  \label{eq:temporal_continuous}
\end{equation}
Performing a linear stability analysis of the disease-free state (i.e., around $p_i(t)=0$) in the quenched mean-field approximation~\cite{Wang2003,Gomez2010}, we obtain
 \begin{equation}
 \dot{p}_i(t) = \sum_j \left[ \lambda A_{ij}(t) - \mu\delta_{ij}\right] p_j(t)\,.
 \label{eq:MarkovContLin}
\end{equation}
We note that this expression is formally equivalent to Eq.~(\ref{eq:contDifferentialComp}). In particular, each row of $P_{ij}$ of Eq.~(\ref{eq:contDifferentialComp}) satisfies Eq.~(\ref{eq:MarkovContLin}). Furthermore, the initial condition $P_{ij}(0) = \delta_{ij}$ guarantees that in varying the row $i$ we consider all vectors of the space basis as initial condition. Every solution $p(T)$ of Eq.~(\ref{eq:MarkovContLin}) can therefore be expressed as a linear combination of the rows of $P(T)$. 
Any fundamental matrix solution of Eq.~(\ref{eq:MarkovContLin}) obeys Eq.~(\ref{eq:contDifferentialComp}) within the framework of Floquet theory of nonautonomous linear systems~\cite{Tian2014}.

The equivalence of the two equations shows that our limit of the discrete-time propagator encodes the dynamics of the continuous process. It is important to note that the limit process leading to Eq.~(\ref{eq:contDifferential}) entails a fundamental change of paradigm on the representation of the network structure and contagion process, where the linear algebraic representation suitable in discrete time turns into a differential geometrical description of the continuous-time flow. While network and spreading dynamics in discrete time are encoded in a multilayer adjacency tensor, the continuous time description proposed in Eq.~(\ref{eq:contDifferentialComp}) rests on a representation of the dynamical process in terms of a manifold whose points are adjacency matrices (or rank-2 tensor in the sense of~\cite{Dedomenico2013}) corresponding to possible network and contagion states. The dynamics of Eq.~(\ref{eq:contDifferentialComp}) is then a curve on such manifold, indicating which adjacency matrices to visit and in which order. In practice, we recover that the contagion process on a discrete temporal network corresponding to an ordered subset of the full multilayer structure of~\cite{Dedomenico2013} becomes in the limit $\Delta t\rightarrow 0$ a spreading on a continuous temporal network represented through a 1-dimensional ordered subset of a tensor field (formally the pull-back on the evolution curve). 
The two frameworks, so far considered independently and mutually exclusive, thus merge coherently through a smooth transition in this novel perspective.

We now turn to solving Eq.~(\ref{eq:contDifferential}) to derive an analytic expression of the infection propagator.
By defining the rescaled transmissibility $\gamma = \lambda/\mu$, we can solve Eq.~(\ref{eq:contDifferential}) in terms of a series in $\mu$~\cite{Blanes2009}
\begin{eqnarray}
P(t) = 1+\sum_{j>0} \mu^j P^{(j)}(t)\,,
\label{eq:P_t_series}
\end{eqnarray}
with $P^{(0)}=1$ and under the assumption that $\gamma$ remains finite around the epidemic threshold for varying recovery rates. The recursion relation from which we derived Eq.~(\ref{eq:contDifferential}) provides the full propagator for $t=T$. Eq.~(\ref{eq:P_t_series}) computed in $T$ therefore yields the infection propagator for the continuous-time adjacency matrix $A(t)$, and is defined by the sum of the following terms
\begin{align}
 P^{(j)}(T) = &\int_0^{T}\dec{x_1}\int_0^{x_1}\dec{x_2}\dots\int_0^{x_{j-1}}\dec{x_j} [\gamma A(x_j)-1]\cdot\nonumber\\
 &\cdot[\gamma A(x_{j-1})-1]\cdots[\gamma A(x_1)-1]\,.
 \label{eq:perturbative_term}
\end{align}
Equations~(\ref{eq:P_t_series}) and (\ref{eq:perturbative_term}) can be put in a compact form by using Dyson's time-ordering operator $\mathcal{T}$~\cite{Dyson1949}. It is defined as $\mathcal{T}A(t_1)A(t_2)=A(t_1)A(t_2)\theta(t_1-t_2) + A(t_2)A(t_1)\theta(t_2-t_1)$, with $\theta$ being Heaviside's step function. The expression of the propagator is thus
\begin{equation}
P(t) = \mathcal{T} \exp \int_0^t \de{x} \left[-\mu + \lambda A(x)\right].
\label{eq:Dyson}
\end{equation}

Eq.~(\ref{eq:Dyson}) represents an explicit general solution for Eq.~(\ref{eq:contDifferential}) that can be computed numerically to arbitrary precision~\cite{Blanes2009}.
The epidemic threshold in the continuous-time limit is then given by $\rho\left(P(T)\right)=1$.


We now discuss a special case where we can recover a closed-form solution of Eq.~(\ref{eq:Dyson}), and thus of the epidemic threshold. We consider continuously evolving temporal networks satisfying the following condition ({\it weak commutation}):
\begin{equation}
 [A(t),\int_0^t\de{x}A(x)]=0, \;\; \forall t\in[0,T]\,,
 \label{eq:commutation_relation}
\end{equation}
i.e. the adjacency matrix at a certain time $A(t)$ commutes with the aggregated matrix up to that time.
In the introduced tensor field formalism, the weak commutation condition represents a constraint on the temporal trajectory, or equivalently,  an equation of motion for $A(t)$. 

Eq.~(\ref{eq:commutation_relation}) implies that the order of factors in Eq.~(\ref{eq:perturbative_term}) no longer matters. Hence, we can simply remove the time-ordering operator $\mathcal{T}$ in Eq.~(\ref{eq:Dyson}), yielding 
\begin{equation}
  P(T) = e^{T\left[-\mu + \lambda \left\langle A \right\rangle  \right]},
\end{equation}
where $\left\langle A \right\rangle = \int_0^{T} \de{t} A(t)/{T}$ is the adjacency matrix averaged over time.
The resulting expression for the epidemic threshold for weakly commuting networks is then
\begin{equation}
 \lambda_c = \frac{\mu}{\rho[\panino{A}]}.
 \label{eq:commutSoglia}
\end{equation}

This closed-form solution proves to be extremely useful as a wide range of network classes satisfies the weak commutation condition of Eq.~(\ref{eq:commutation_relation}). 
An important class  is constituted by annealed networks~\cite{Boguna2009,PastorSatorras2001,Castellano2010}. In the absence of dynamical correlations, the annealed regime leads to $\langle [A(x),A(y)]\rangle = 0$, as the time ordering of contacts becomes irrelevant. Eq.~(\ref{eq:commutation_relation}) can thus be reinterpreted as $\langle[A(t),A(x)]\rangle_x = 0$, where the average is carried out over $x\in\left[0,t\right)$. For long enough $t$, $\int_0^{t} \de{x} A(x)/t$ approximates well the expected adjacency matrix $\panino{A}$ of the annealed model, leading  the annealed regime to satisfy Eq.~(\ref{eq:commutSoglia}).  This result  thus provides  an alternative mathematical framework for the conceptual interpretation of annealed networks in terms of weak commutation. Originally introduced to describe disorder on  quenched networks~\cite{Gil2005,Stauffer2005}, annealed networks were mathematically described in probabilistic terms, with the probability of establishing a contact depending on the degree distribution $P(k)$ and the two-node degree correlations $P(k'|k)$~\cite{Boguna2009}. Here we show that temporal networks whose adjacency matrix $A(t)$ asymptotically commutes with the expected adjacency matrix are found to be in the annealed regime. 

Eq.~(\ref{eq:commutSoglia}) can also be used to test the limits of the time scale separation approach, by considering a generic temporal network not satisfying the weak commutation condition. If $\mu$ is small, we can truncate the series of the infection propagator (Eq.~(\ref{eq:P_t_series})) at the first order, $P = 1+\mu P^{(1)} + \mca{O}(\mu^2)$, where $P^{(1)}(T)=T[\gamma\panino{A}-1]$, to recover indeed Eq.~(\ref{eq:commutSoglia}). The truncation thus provides a mathematical expression of the range of validity of the time-separation scheme for spreading processes on temporal networks, since temporal correlations can be disregarded when the network evolves much faster than the spreading process.

Extending the result of the annealed networks, we show that the weak commutation condition also holds for networks whose expected adjacency matrix depends on time as a scalar function (instead of being constant as in the annealed case), $\panino{A(t)} = c(t)\panino{A(0)}$. Also in this case we have $\langle [A(x),A(y)]\rangle = 0$, so that the same treatment performed for annealed networks applies. Examples are provided by global trends in activation patterns, as  often considered in infectious disease epidemiology to model seasonal variations of human contact patterns (e.g. due to the school calendar)~\cite{London1973}.

When the time scale separation approach is not applicable, we find another class of weakly commuting temporal networks  that are used as a paradigmatic network example for the study of contagion processes occurring on the same time scale of contacts evolution~--~the activity-driven model~\cite{Perra2012}.
It considers heterogeneous populations where each node $i$ activates according to an activity rate $a_i$, drawn from a distribution $f(a)$. When active, the node establishes $m$ connections with randomly chosen nodes lasting a short time $\delta$ ($\delta\ll 1/a_i$). Since the dynamics lacks time correlations, the weak commutation condition holds, and the epidemic threshold can be computed from Eq.~(\ref{eq:commutSoglia}). In the limit of large network size, it is possible to write the average adjacency matrix as $\panino{A}_{ij} = \frac{m\delta}{N}(a_i+a_j)+\mca{O}(\frac{1}{N^2})$. Through row operations we find that the matrix has $\mbox{rank}(\panino{A})=2$, and thus only two non-zero eigenvalues, $\alpha,\sigma$, with $\alpha>\sigma$. We compute them through the traces of $\panino{A}$ ($\mbox{tr}\!\left[\panino{A}\right]=\alpha+\sigma$ and $\mbox{tr}[\panino{A}^2]=\alpha^2+\sigma^2$) to obtain the expression of $\rho[\panino{A}]$ for Eq.~(\ref{eq:commutSoglia}): $\rho[\panino{A}] = \alpha = m\delta \left( \panino{a}+\sqrt{\panino{a^2}}\right)$. The epidemic threshold becomes
\begin{equation}
 \lambda_c \delta = \frac{\mu}{m\left( \panino{a}+\sqrt{\panino{a^2}}\right)},
\end{equation}
yielding the same result of Ref.~\cite{Perra2012}, provided here that the transmission rate $\lambda$ is multiplied by $\delta$ to make it a probability, as in~\cite{Perra2012}.

Finally, we verify that for the trivial example of static networks, with an adjacency matrix  constant in time,  Eq.~(\ref{eq:commutSoglia}) reduces immediately to the result of Refs.~\cite{Wang2003,Gomez2010}.

 
\begin{figure}
\begin{center}
 \includegraphics[width=0.65\textwidth]{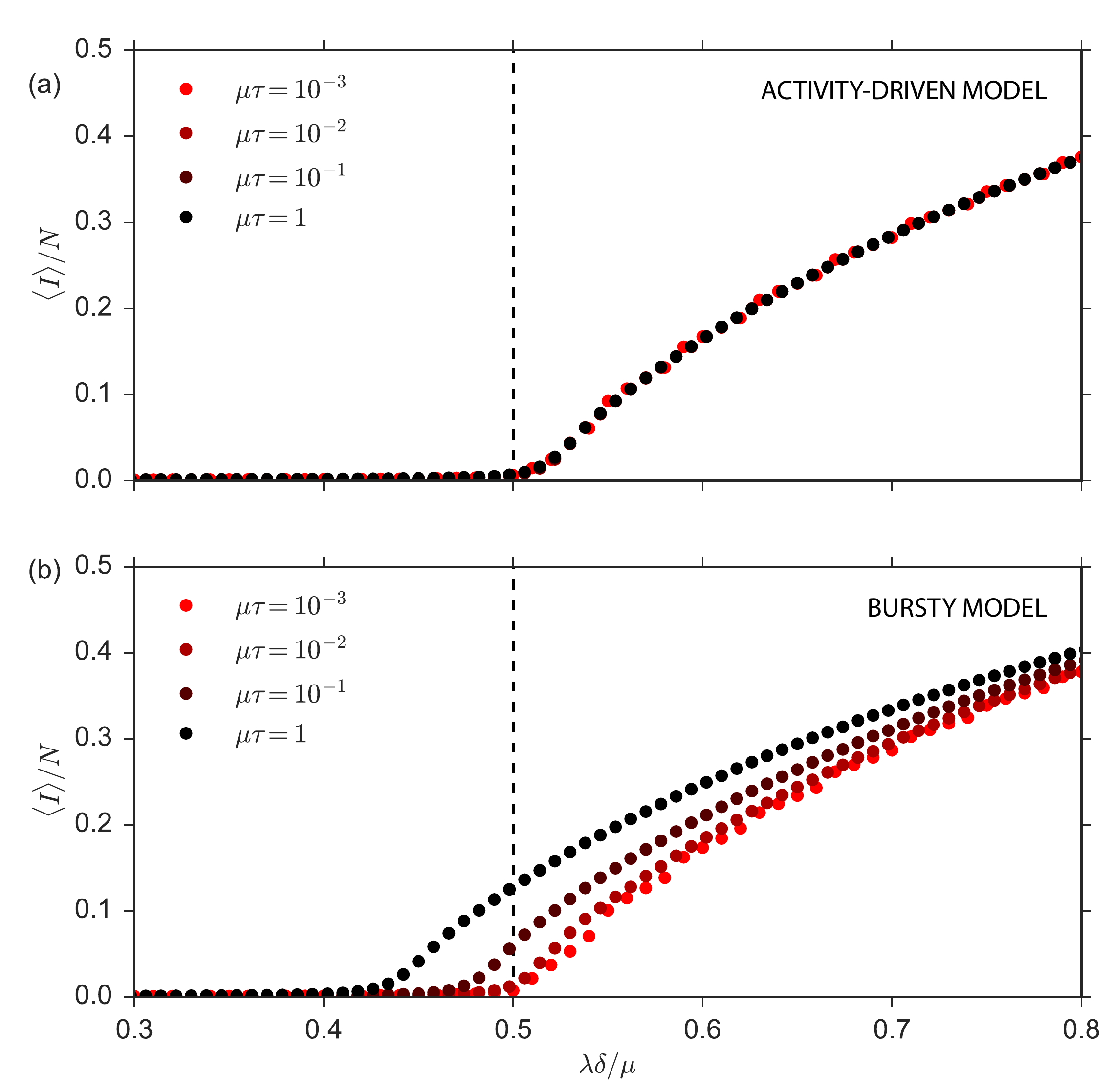}
\caption{\label{fig:plot} Performance of the infection propagator estimate of the epidemic threshold in the continuous-time limit under the weak commutation approximation (Eq.~(\ref{eq:commutSoglia})). Panels report the average simulated endemic prevalence as a function of $\lambda\delta/\mu$ for the activity-driven model (a) and  the bursty model (b). Different colors refer to explored values of the  recovery rate $\mu$. The vertical dashed line is the prediction for the critical transmissibility provided by Eq.~(\ref{eq:commutSoglia}).
}
\end{center}
\end{figure}

We now validate our analytical prediction against numerical simulations on two synthetic models. The first is the activity-driven model with activation rate $a_i=a$, $m=1$, and average inter-activation time $\tau=1/a=1$, fixed as the time unit of the simulations. The transmission parameter is the probability upon contact $\lambda\delta$ and the model is implemented in continuous time. 
The second model is based on a bursty inter-activation time distribution $P(\Delta t)\sim (\epsilon+\Delta t)^{-\beta}$~\cite{Rocha2013}, with $\beta=2.5$ and $\epsilon$ tuned to obtain the same average inter-activation time as before, $\tau=1$. We simulate a SIS spreading process on the two networks with four different recovery rates,  $\mu\in\{10^{-3},10^{-2},10^{-1},1\}$, i.e. ranging from a value that is 3 orders of magnitude larger than the time scale $\tau$ of the networks (slow disease), to a value equal to $\tau$ (fast disease). We compute the average simulated endemic prevalence for specific values of $\lambda,\mu$ using the quasi-stationary method~\cite{Ferreira2012} and compare the threshold computed with Eq.~(\ref{eq:commutSoglia}) with the simulated critical transition from extinction to endemic state. As expected, we find Eq.~(\ref{eq:commutSoglia}) to hold for the activity-driven model at all time scales of the epidemic process (Fig.~\ref{fig:plot}), as the network lacks temporal correlations. The agreement with the transition observed in the bursty model however is recovered only for slow diseases, as at those time scales the network is found in the annealed regime. When network and disease time scales become comparable, the weakly commuting approximation of Eq.~(\ref{eq:commutSoglia}) no longer holds, as burstiness results in dynamical correlations in the network evolution~\cite{Rocha2013}. 

Our theory offers a novel mathematical framework that rigorously connects discrete-time and continuous-time critical behaviors of spreading processes on temporal networks. It uncovers a coherent transition from an adjacency tensor to a tensor field resulting from a limit performed on the structural representation of the network and contagion process. We derive an analytic expression of the infection propagator in the general case that assumes a closed-form solution in the introduced class of weakly commuting networks. This allows us to provide a rigorous mathematical interpretation of annealed networks, encompassing the different definitions historically introduced in the literature.
This work also provides the basis for important theoretical extensions, assessing, for example, the impact of bursty activation patterns or of the adaptive dynamics in response to the circulating epidemic. Finally, our approach offers a tool for applicative studies on the estimation of the vulnerability of temporal networks to contagion processes in many real-world scenarios, for which the discrete-time assumption would be inadequate.

\vspace{2cm}
\noindent
{\bf Acknowledgments}
\noindent
We thank Luca Ferreri and Mason Porter for fruitful discussions. This work is partially sponsored by the EC-Health contract no. 278433 (PREDEMICS) and the  ANR contract no. ANR-12-MONU-0018 (HARMSFLU) to VC; the EC-ANIHWA contract no. ANR-13-ANWA-0007-03 (LIVEepi) to EV, CP, VC.

\newpage










\end{document}